# Optimizing vaccine distribution networks in low and middle-income countries

Yuwen Yang, Hoda Bidkhori, Jayant Rajgopal*

Department of Industrial Engineering

University of Pittsburgh

Pittsburgh, PA 15261

*Corresponding author:

1032 Benedum Hall, Department of Industrial Engineering, University of Pittsburgh, Pittsburgh, PA 15261 (*rajgopal@pitt.edu*)

# Optimizing vaccine distribution networks in low and middle-income countries

## Abstract

Vaccination has been proven to be the most effective method to prevent infectious diseases. However, there are still millions of children in low and middle-income countries who are not covered by routine vaccines and remain at risk. The World Health Organization – Expanded Programme on Immunization (WHO-EPI) was designed to provide universal childhood vaccine access for children across the world and in this work, we address the design of the distribution network for WHO-EPI vaccines. In particular, we formulate the network design problem as a mixed integer program (MIP) and present a new algorithm for typical problems that are too large to be solved using commercial MIP software. We test the algorithm using data derived from four different countries in sub-Saharan Africa and show that the algorithm is able to obtain high-quality solutions for even the largest problems within a few minutes.

**Keywords**: Vaccines, Network design, Supply chains, Mixed integer programming

# 1. Introduction

Vaccination has been proven to be the most effective method to prevent illness, disability and death from infections. It is estimated that 2 to 3 million deaths are averted each year because of vaccines [1] and over the years, significant levels of coverage have been achieved. However, it is estimated that an additional 1.5 million deaths could be avoided annually if global vaccination coverage could improve further, and that even in the 21$^{st}$ century there are still almost 20 million infants worldwide who lack access to routine immunization services and remain at risk for vaccine-preventable diseases [2]. This problem is especially pronounced in low and middle-income countries (LMICs), where some of the contributors to the problem include high costs, competing health priorities, lack of resources, inadequate infrastructure, poor monitoring and supervision, rigid distribution structures, and even religious reasons [3–5].

The World Health Organization (WHO) established the Expanded Programme on Immunization (EPI) in 1974 with the goal of providing universal access to all important vaccines for all children [6]. The program was further expanded with the formation of the Global Alliance for Vaccines and Immunization (Gavi) in 2000 to accelerate access to new vaccines in the poorest countries. EPI and Gavi together have successful contributed to saving millions of lives worldwide by reducing mortality and even largely eliminating some diseases like polio and measles [7,8].

With the help of international organizations and new technological developments, many vaccines can now be obtained at low cost and in mass quantities. However, shipping, storing and delivering vaccines in a cost-efficient fashion while ensuring that vaccines are reliably available to end-users remains a major challenge. In particular, in many LMICs vaccines are usually distributed via a hierarchical legacy medical network, with locations and shipping routes of this network often determined by political boundaries and history. The overarching goal is to ensure that every child has access to vaccines, and along with this, in most LMICs the objective is to design a system that can be operated without the need for sophisticated logistics personnel and at minimum cost.

This fact motivates our study to propose an improved vaccine distribution network. As an alternative to the current structure, the network for vaccine distribution could be separated from the current legacy health network, while using some appropriate subset of these facilities and with vaccine flows along routes that differ from the current ones. However, the operation of the network must not deviate from established WHO guidelines and needs to be simple because of the relative lack of sophisticated vaccine management abilities in LMICs.

In this paper we develop a mixed integer programming (MIP) model that optimizes the design of the distribution network. The model allows for a more appropriate design than current networks while following the WHO guidelines for operational simplicity. The resulting formulation for a national network is too large to solve optimally using standard commercial software and we develop a novel algorithm that solves a sequence of increasingly larger MIP problems. To maintain tractability, the approach uses insights into the problem structure and principles from cluster analysis to limit the size of each MIP in the sequence. In order to study the performance of the algorithm, numerical tests using data derived from several different countries in sub-Saharan Africa are conducted. Comparisons with the optimal solution when one is available indicate that the algorithm works very well with solution times that scale up in a roughly linear fashion.

## 2. Problem development and literature review

In most LMICs, vaccines are distributed via a four-tier hierarchical legacy medical network such as the one depicted in Figure 1. Typically, EPI vaccines are purchased in bulk and shipped in by air once or twice a year, then stored in a national distribution center in the capital (or other large city). Required vaccine volumes are transported every three months to regional distribution centers using a specialized vehicle such as a large cold truck. Each regional distribution center delivers vaccines to its surrounding district centers every month using a smaller cold truck or more commonly, 4×4 trucks with cold storage boxes. Finally, the vaccines are transported from district centers in a vaccine carrier/cooler using locally available means of transportation such as trucks, cars, motorbikes, bicycles, boats, or sometimes even by foot, to local clinics where infants, children and pregnant women come to receive vaccinations. This last step is typically, a "pull" operation with monthly pickup by the clinic. A characteristic of EPI vaccines is that they must be stored/transported while maintaining appropriate temperatures (2° to 8°C), so that this vaccine distribution chain is often referred to as being a cold chain.

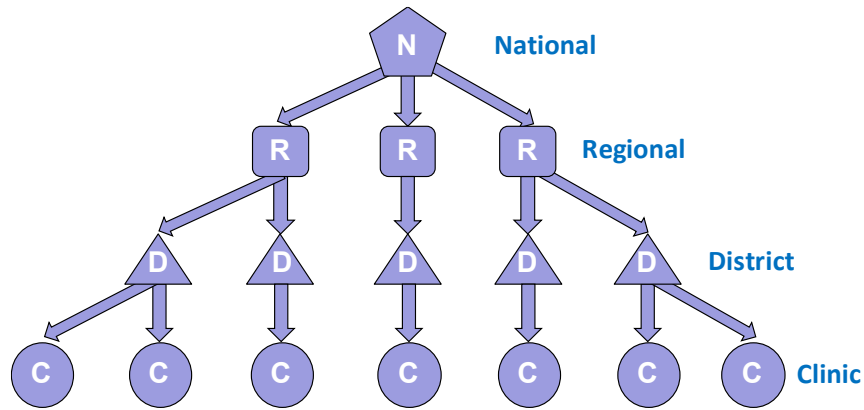

**Figure 1.** A typical tour-tier vaccine distribution network

To develop an optimal vaccine distribution network design and operational policies for a country that also follow WHO guidelines, we separate this from the existing legacy medical supply chain of which it is typically a component, and model the cold chain independently. Our objective is to minimize the overall cost of transportation, facilities and storage over the whole network, while guaranteeing universal access and following WHO operational guidelines. In the model, vaccines flow from the national center (source node) to clinics (sink nodes), usually via one or more intermediate hubs (transshipment nodes). Although multiple (usually 6 to 8) vaccines are handled in the cold chain, transportation and storage capacities are only affected by the overall space required. Therefore, we only consider the total volume of vaccines shipped or stored. Hub locations are chosen from the current locations of legacy intermediate nodes (regional or district center), and while we retain the choices of monthly or quarterly replenishments as per WHO guidelines, we allow a hub to freely select either option. The model determines the clinics assigned to each hub, the national center-to-hub and hub-to-hub connections, the actual vaccine flows on these connections, and the types of storage and transportation devices to be deployed at each location and for each flow.

We make the following assumptions to model the EPI vaccine network:
1) The network should be capable of meeting all demand that can arise at each clinic, and this demand is determined by the estimated population that the particular clinic serves.
2) The locations of the clinics and the national distribution center are fixed but we can choose hub distribution centers freely from the current set of regional and district centers.

3) Each clinic is assigned to a hub for its vaccines (although clinics close to the national center could be directly supplied by it) and replenished once a month. Each hub is supplied by the national center or by another hub.
4) The national center is the root node of the network and all other nodes (hubs and clinics) have exactly one inbound arc.
5) As per WHO guidelines, a hub is replenished either quarterly or monthly.
6) Every open facility has an appropriately sized storage device, to be selected from the WHO's pre-qualified list of devices.
7) As per WHO guidelines, there is a 25% safety buffer at each clinic location so that the total demand volume is inflated by this factor.

We now discuss these assumptions briefly. First, universal access is the goal of the WHO and our model's constraints explicitly capture this. Second, our model is based on using the existing facilities for hubs as opposed to building new ones. Third, in most LMICs, operational simplicity is a driving requirement because resources are very constrained and it can be a challenge to find qualified logisticians and trained personnel who can deal with multiple suppliers and different types of equipment. We therefore retain the current approach of restricting each facility location to having a single supplier, and a single type of storage device that is selected from the WHO's pre-qualified list [9]. Finally, for the same reason of operational ease, we do not attempt to determine optimal safety buffers at clinics or reorder points by location; hub locations are restricted to one of two replenishment intervals (monthly or quarterly) and all clinics have the same 25% buffer inventory levels as per WHO guidelines.

Several variations of this class of network design problems have been addressed by the operations research community, including the *p*-median problem, the uncapacitated and capacitated facility location problems [10–12], and extensions to include transportation cost [13]. The facility location problem is often combined with the vehicle routing problem and typically using heuristics [14]. It has also been extended with consideration of risk pooling [15] and facility failures [16], with Lagrangian relaxation being a common solution strategy [17]. Klose and Drexl [18] reviewed several facility location models for distribution system design.

As an extension to these facility location problems, the hub selection problem considers the situation where one or more nodes are designated as facilities that serve as consolidation, switching or transshipment points and connect to origin/destination nodes. It has received attention in applications ranging from airlines and emergency services to intermodal logistics and postal delivery services. Interested readers are referred to the surveys on hub location problems presented in [19,20]. Models for these problems address a variety of objectives and combinations of various problem environments [19]. These include the domain of the hub nodes (all network nodes, discrete subset or anywhere along a continuous plane), the number of nodes to be designated as hubs (pre-specified vs. unspecified), hub capacity (limited vs. unlimited), cost of locating hubs (none, fixed or variable), allocation of non-hub nodes to hubs (single vs. multiple) and allocation costs (none, fixed or variable). In particular, our formulation chooses an unspecified number of hubs from a discrete subset of capacitated hub locations that incur fixed costs, and each non-hub locations is assigned to a single hub with allocation costs that are exogenous to the model. In studying the literature for exact algorithms to find the optimum solution to models similar to ours (capacitated, single allocation, $p$-hubs; albeit with differing objectives), approaches include bi-criteria integer linear programming [21], mixed integer programming [22,23], generalized Benders' decomposition [24], and fuzzy integer linear programming [25]. The largest problem size solved is reported as having "up to 10,000 integer variables" [24]. Given the limits to the size of the problems that can be solved optimally, many others have resorted either to metaheuristics [26], heuristics based on Lagrangean relaxation or Benders' decomposition [27,28] or heuristics designed for specific formulations [29].

A unique aspect of our model is that replenishment frequencies as well as capacities at hubs and for transportation along arcs are limited to multiple discrete options. This greatly increases the number of binary variables to well over 10,000 even for medium sized problems, and while most of the prior work looks at under 20 hubs, our formulation for an entire country has a number of hubs as well as a total number of binary variables that are an order of magnitude higher. Thus, solving full-country problems using an exact approach is not a viable option. Results from our initial experiments with using Lagrangean relaxation were also not encouraging. Therefore, we develop a heuristic approach that is designed for the application domain and the specific class of problems that we study. These and related computational issues are discussed in detail in Sections 3 and 4.

In terms of work specific to vaccine distribution networks, Chen et al. [30] were the first to model the network in 2014 as a planning model to maximize the number of children being fully immunized under current network capacity; they then extended it to the case where capacity expansion is allowed. However, this work addresses operations in an existing network as opposed to its design. In 2016, Lim [31] proposed a model to design a minimum cost vaccine distribution network, and utilized an evolutionary strategy to solve this problem. His model assumes that deliveries to hubs are coordinated and done using vehicle loops and fixes the storage devices; thus multiple trips might have to be made along a route if the volume cannot be handled in a single trip.

In Section 2.1 we present a mixed integer programming model that draws on the initial work by Lim [31] but allows for flexibility in replenishment, allows storage devices to be selected in the required size, does not require delivery coordination during replenishment and ensures that all deliveries to a node are made in a single trip as is typically the case in practice. We also develop a mathematical programming based heuristic to address larger problems that cannot be solved via standard commercial software.

## 2.1 Formulation

We now develop our model formulation.

*Index sets*:

$N$: National Distribution Center = $\{0\}$

$H$: Potential Hub Distribution Centers = $\{1, 2 \ldots h\}$

$C$: Local Clinics = $\{h+1 \ldots n\}$, where $n = |H| + |C|$

$V$: Vertices: $N \cup H \cup C$

$A$: Arcs: $(i, j) | i \in N \cup H, j \in H \cup C; i \neq j$

$T$: Transportation Vehicles

$R$: Storage Devices

$F$: Replenishment Frequency: {Quarterly (=0), Monthly (=1)}

*Parameters*:

$c_{ijt}^P$: Transportation cost per km of vehicle type $t$ between locations $i$ and $j$ ; $(i,j) \in A; t \in T$

$c_{jr}^S$: Annual facility cost when facility $j$ is open and uses storage device $r$; $r \in R$

$p_t^P$: Transportation capacity per trip of vehicle $t$; $t \in T$

$p_r^S$: Storage capacity of device $r$; $r \in R$

$g_f$: Annual number of replenishments; $f \in F$ ($g_f = 4$ if $f = 0$; $g_f = 12$ if $f = 1$)

$d_{ij}$: Driving distance (km) between location $i$ and location $j$; $(i, j) \in A$

$b_j$: Annual demand volume at location $j$, $j \in C$

*Variables*:

$U_{ijtf} \in \{0,1\}$: 1 if vaccines flow from location $i$ to location $j$ using vehicle type $t \in T$ and with replenishment frequency $f \in F$, 0 otherwise; $i \in N \cup H, j \in H \cup C$

$W_{irf} \in \{0,1\}$: 1 if hub location $i \in H$ is open and uses storage device of type $r \in R$ and replenishment frequency $f \in F$, 0 otherwise

$X_{ij}$: Annual flow (volume) of vaccines from location $i$ to location $j$; $i \in N \cup H, j \in H \cup C$

The mixed integer program for designing the optimal network may then be formulated as follows:

**Program MIP-1**

$$\text{Min} \sum_{j \in H} \sum_{r \in R} \sum_{f \in F} c_{jr}^S W_{jrf} + \sum_{(i,j) \in A} \sum_{t \in T} \sum_{f \in F} 2 c_{ijt}^P g_f d_{ij} U_{ijtf} \tag{1}$$

*subject to*

$$\sum_{r \in R} \sum_{f \in F} W_{jrf} \leq 1 \qquad j \in H \tag{2}$$

$$\sum_{i \in N \cup H} \sum_{t \in T} U_{ijt1} = 1 \qquad j \in C \tag{3}$$

$$\sum_{i \in N \cup H} \sum_{t \in T} \sum_{f \in F} U_{ijtf} \leq 1 \qquad j \in H \tag{4}$$

$$\sum_{r \in R} W_{jrf} - \sum_{i \in N \cup H} \sum_{t \in T} U_{ijtf} = 0 \qquad j \in H, f \in F \tag{5}$$

$$\sum_{i \in N \cup H} X_{ij} - \sum_{k \in H \cup C} X_{jk} = 0 \qquad j \in H \tag{6}$$

$$\sum_{i \in N \cup H} X_{ij} = b_j \qquad j \in C \tag{7}$$

$$\sum_{r \in R} \sum_{f \in F} p_r^S W_{jrf} - (1/g_f) \sum_{i \in N \cup H} X_{ij} \geq 0 \qquad j \in H \qquad (8)$$

$$\sum_{t \in T} \sum_{f \in F} p_t^P U_{ijtf} - (1/g_f) X_{ij} \geq 0 \qquad i \in N \cup H, j \in H \cup C \qquad (9)$$

$$X_{ij} \geq 0 \qquad i \in N \cup H, j \in H \cup C \qquad (10)$$

$$W_{jrf} \in \{0,1\} \qquad j \in H, r \in R, f \in F \qquad (11)$$

$$U_{ijtf} \in \{0,1\} \qquad i \in N \cup H, j \in H \cup C, t \in T, f \in F \qquad (12)$$

The objective function (1) has two components: annual hub facility costs and total annual round-trip transportation costs. Constraints (2) ensure that every open hub $j$ has a single replenishment frequency and a single type of storage device, while constraints (3) ensure that each clinic has exactly one inflow and a monthly replenishment frequency. Constraints (4) ensure that each hub has at most one inflow with unique associated replenishment frequency and transport device. Constraints (5) ensure there is no flow associated with a hub that is not open and constraints (6) and (7) are standard flow balance equations at hubs and clinics. Constraints (8) ensure that in each hub there is a sufficiently large storage device to store the vaccines required within each replenishment interval. Finally, constraints (9) ensure that a transportation mode with sufficient capacity is selected to carry the required volume of vaccines for replenishment.

## 2.2 Limitations with solving MIP-1

To explore the solution of the model described by MIP-1 we tested it with a standard commercial solver using data derived from the EPI networks in four different countries in sub-Saharan Africa; specifics on the data, as well as the hardware and software used are discussed in Section 4, where we describe our numerical experiments in full detail.

Unfortunately, none of the models for these countries could be directly solved using off-the-shelf commercial software. To further explore the limits of the problem size that could be solved using a standard solver we also experimented with subsets of the data from each country. That is, we considered successively larger problems: first, with the national center along with a single region (based on how a region is currently defined in the country), then problems with a combination of two regions, three regions, etc. In general, the difficulty associated with a particular problem depends on several factors including the total number of nodes and potential hubs in the problem, as well as the population distribution, transportation cost, and storage cost across the network. Unfortunately, despite extensive computational experimentation it was impossible to pinpoint the limiting characteristics of a tractable problem or establish any clear monotonicity, because of the interrelationships between the problem parameters. Our numerical tests are discussed in more detail in Section 4, but as a general rule of thumb, we found that most problems with over 200 to 250 nodes and over 15 to 20 potential hubs are impossible to solve directly. Given that in the network for an entire country these limits are almost always exceeded, there is clearly a need for good algorithm if one aims to design an optimal network for the country.

A key fact that makes MIP-1 hard to solve is that the model has a large number of 0-1 decision variables. For example, if we can choose from three types of transport vehicles and four types of storage devices (i.e., $\{T\}=3$, $|R|=4$), Table 1 illustrates the number of decision variables in MIP-1. Thus, in order to solve the full problem for one our instances with 685 nodes and 41 candidate hubs, we end up with 168,838 integer variables. Even for a typical 100-node, mid-sized problem with 15 candidate hubs, the number of binary decision variables is close to 10,000.

**Table 1.** Number of decision variables in MIP for a problem with *n* nodes and *h* potential hubs

| **Decision Variables** | $W_{jrf}$ | $U_{ijtf}$ | $X_{ij}$ |
|---|---|---|---|
| Type | Integer | Integer | Continuous |
| Number | $8h$ | $6hn$ | $hn$ |

In the next section, we propose a sequential MIP-based disaggregation-and-merging algorithm that divides the problem on the entire graph for the distribution network into several subproblems on smaller subgraphs that can be solved with relative ease. The algorithm then intelligently merges the subgraphs together sequentially to obtain a solution to the whole graph. We present numerical comparisons in Section 4 and show that the algorithm is able to yield good solutions for even the largest problems.

## 3. A disaggregation-and-merging algorithm

Our algorithm is motivated by the observation that in a large network, the optimal subnetwork structures in regions that are relatively far apart will tend to be independent of each other. For example, the characteristics of a local clinic are unlikely to have any influence on the network structure in locations that are far away, and if a hub is added or removed at some distant location it is unlikely to affect the clinic's supply. The same is also true of a hub that is distant from some other hub whose disposition is changed. Therefore, for larger problems we propose a divide-and-conquer approach where we first divide the whole network into portions that each yield smaller problems that can be solved independently with relative ease. We then systematically merge these smaller problems and solve a sequence of increasingly larger problems. Each of these is formulated using MIP-1, but with the key difference that parts of the structure are fixed based on the optimal solutions to the smaller problems as well as the spatial relationship between the current subnetwork and the new portion being added on. To clarify our approach, we first provide an overview of the method and then provide all of the details.

We start by dividing the entire network into $P$ smaller subnetworks. One could always use existing political boundaries as a natural disaggregation of the network, i.e., each region or province or state of the country is an "independent" network; larger existing regions could be split into smaller ones. Alternatively, we could apply a clustering algorithm to determine them. Although the average cluster size will be smaller as we form more clusters, the number of clusters that would give us problems that are small enough to yield a tractable version of MIP-1 is problem dependent, so that it is difficult to prescribe a general value for $P$ in advance. We therefore chose to use *hierarchical clustering* rather than a simpler method such as *K-means clustering*, whereby we can continue to disaggregate the network until each region is small enough for a standard MIP solver to handle; the interested reader is referred to [32] for more details on hierarchical clustering.

Once the independent regions are created we start with the one contains the national center and optimize its structure via MIP-1 to obtain an initial subnetwork. We now pick a neighboring region to merge with this subnetwork, formulate MIP-1 for the combined set of nodes and solve this (larger) consolidated problem to get a new subnetwork structure with both regions. This process continues until all of the independent regions have been merged into our network. While we will specify details on how each step is executed, the critical thing to note is that at each successive iteration we handle a larger collection of nodes, and therefore have to solve a larger problem. Clearly, the effort required at each stage has to be reasonable; otherwise we are defeating the purpose of the original disaggregation! To ensure that this is the case, we refer to our initial observations on the motivation for this approach, and at each iteration we fix a portion of the current subnetwork, so that we are only using variables associated with a subset of all the nodes corresponding to the current iteration's network. This is done by retaining the locally optimal structure for portions of the subnetwork while allowing for changes in others. In addition, we also use a "shrinking" scheme whereby some of the nodes are aggregated and replaced by a single dummy node so as to further reduce the size of the problem being solved.

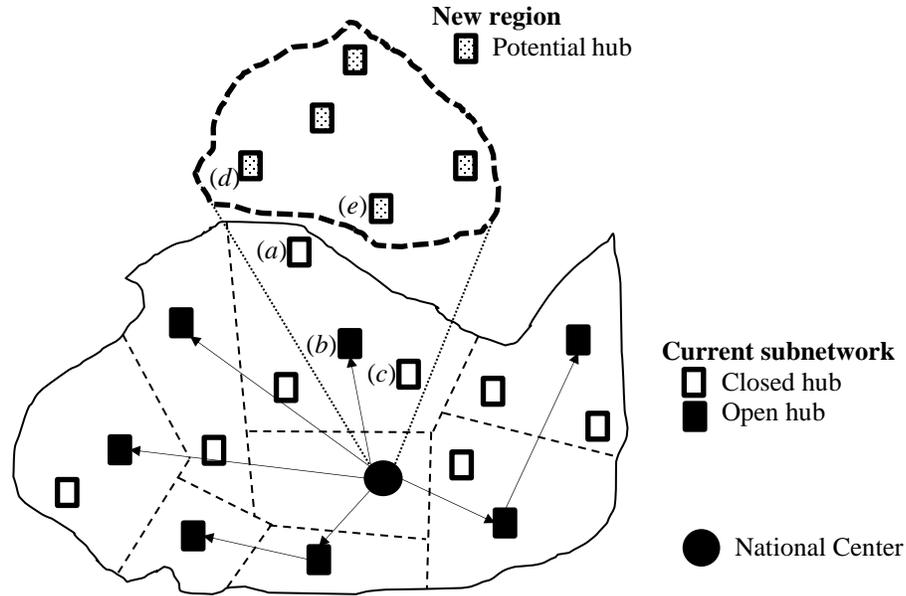

**Figure 2**. Hub classification during consolidation

To decide upon how to fix parts of the structure at each iteration, we classify all hubs for the current network into three categories. We overview these categories here using the example shown in Figure 2; more mathematically precise definitions are provided when the algorithm is detailed. The lower part of Figure 2 shows the subnetwork we currently have. The seven solid squares represent open hubs while the empty squares denote potential hub locations that were not selected for opening. The dashed lines within the area covered by this subnetwork divide it into eight discrete sections corresponding to the national center and the seven open hubs, and clinics within each section (not shown) are supplied by the corresponding open hub (or national center). Note that some open hubs are supplied directly by the national center while others receive their vaccines from some other open hub. The upper portion of Figure 2 represents the new, neighboring region that we wish to merge with the current subnetwork, along with five potential hub locations within it.

- The first category, which we call *Critical Hubs*, are hub locations "close to" the boundary between the current subnetwork and the new region being merged. The dispositions of these hubs are likelier to change after merger; currently open hubs might close and vice-versa. Clinic assignments to a hub might also change as new hubs might be introduced at geographically proximal locations. In our illustration, nodes (*a*), (*d*) and (*e*) might be critical hubs.
- The second category, which we call *Intermediate Hubs*, are hub locations in the existing subnetwork that are in some sense "in between" the national center and hubs in the new region that is being merged with the existing subnetwork (e.g., hubs labeled (*a*), (*b*) and (*c*) in Figure 2). Since they are en route from the center to a possible hub in the new region, such hubs could potentially serve as intermediate transshipment points (while continuing to serve their current clinics if they are currently open). Thus their storage requirements could be larger after merging and/or their replenishment frequencies could possibly change.
- The third category of hub locations, which we call *Non-critical Hubs* can be considered as "independent" during merger. These are locations that are not near the common boundaries or en route to a potential new hub and we fix their dispositions (open or closed); for the open hubs, their clinic assignments, hub-to-hub connections, storage devices, transportation routes and frequencies are also fixed. In Figure 2 all non-labeled nodes might be non-critical hubs.

Lastly, we overview the process to sequence regions for merger. Once an initial set of regions is formed, we first apply MIP-1 just to the nodes in a region along with the national center, and solve a sub-problem for each region to obtain a locally optimal structure for each. At each iteration, we will choose the region for merger as one with minimum cluster distance between it and some region that has already been merged into the consolidated network. Here, we define the cluster distance as the minimum distance between two points that are in different clusters; we found this worked best among the common measures of cluster distance.

We are now ready to outline the steps in our algorithm.

### 3.1 Algorithm 1

**STEP 1**: Disaggregation

Consider the directed graph $G$ of nodes indexed in $V$ and arcs indexed in $A$. Divide the set of potential hubs $H$ into $P$ mutually exclusive subsets $H_1, H_2, ..., H_P$ using a clustering algorithm or heuristically. If using a clustering algorithm, to determine whether a cluster of potential hubs indexed in $H_p$ is small enough, define set $C_p$ to be the set of clinics whose closest hub nodes are in $H_p$ and define set $Q_p = H_p \cup C_p \cup \{0\}$ as the complete node set of the region defined by hub nodes in cluster $H_p$. While there is no obvious way to prescribe the value of $P$ in advance, it is important for the number of nodes in each set $Q_p$ to be small enough that MIP-1 on the subgraph generated by nodes in this set is readily solved. Therefore, define a suitable number $M$ (we suggest a value under 200 based upon our computational experiments) and check whether $|Q_p| \leq M$ for all $p$; if not, further divide the corresponding cluster $H_p$, again by using existing political boundaries (such as districts within the region), a clustering algorithm, or heuristically based on the distribution of nodes within the cluster. Continue until the number of nodes in each of $H_1, H_2, ..., H_P$ is no larger than $M$. Note that depending on the demographic characteristics, the number of nodes in each cluster $H_p$ could be quite different. Define $H'$ as the set of potential hubs that are currently *not* merged into the consolidated network; thus, at the end of the **STEP 1**, $H' = H = \{H_1, H_2, ..., H_P\}$.

**STEP 2**: Sub-problem solution and initialization

Consider the subgraph $G[Q_p]$ that is induced by nodes in $Q_p$. For each $p \in \{1, 2, ... P\}$ formulate and solve MIP-1 on subgraph $G[Q_p]$ to meet demand optimally at all clinic locations in set $Q_p$.

Denoting the cluster distance (i.e., distance of the nearest pair of nodes in different clusters) between clusters $H_p$ and $H_q$ as $D(H_p, H_q)$, compute

$$p^* \in Argmin_{p \in 1,2,...,P}\ D(H_p, N)$$

where cluster $N = \{0\}$ is an artificial cluster with just the national center in it. Thus cluster $H_p$ has the smallest cluster distance to $\{0\}$.

To start the iterative process set $k = 0$, define $I^0 = Q_{p^*} = H_p \cup C_p \cup \{0\}$ as the index set of all nodes in the initial subnetwork, with corresponding subgraph $G^0 = G[I^0]$.

Update $H' \leftarrow H' \setminus \{H_{p^*}\}$.

**STEP 3**: Subset Selection

Set $k = k + 1$ and compute $p^* \in Argmin_{p|H_p \in H'} \{D(H_q, H_p) | H_q \notin H'\}$

Here $q$ and $p$ correspond respectively, to clusters that have and have not yet been merged into the consolidated network, and among the clusters not yet merged $p^*$ has the smallest cluster distance to a cluster that has already been merged. Define $I^k = H_{p^*} \cup C_{p^*} \cup I^{k-1}$ as the complete node set for the consolidated network at iteration $k$. Define the graph $G^k = G[I^k]$ and update $H' \leftarrow H' \setminus \{H_{p^*}\}$.

**STEP 4**: Classification

Compute $d_{max} = Max_{i,j \in H_{p^*}} d_{ij}$. That is, $d_{max}$ is the largest distance between any pair of locations within the new cluster of potential hubs that was just merged. Let $Conv_{p^*}$ be the convex hull of $\{0\}$ and all hub nodes in $H_{p^*}$: $Conv_{p^*} = Conv(H_{p^*} \cup \{0\})$, and define a positive real number $\alpha \in (0,1)$. Classify the hubs in $I^k$ into three categories as follow:

a. Critical Hubs ($H^C$): Identify all pairs of nodes $(i, j)$, such that $i \in I^{k-1} \cap H$ and $j \in H_{p^*}$, with $d_{ij} < \alpha d_{max}$, and define $i$ and $j$ as critical hubs. That is, we consider all potential hub pairs with one from the previous consolidated region and one from the new region, and define the two as being critical if they are separated by less than some fraction of the maximum distance between two hub locations in the region being merged. Larger values for $\alpha$ result in more hubs being identified as critical so that the structure of the consolidated network is more flexible, but the associated model formulation is also more difficult to solve. Conversely, when $\alpha$ is smaller, the consolidated problem is easier to solve but a larger portion of the network is fixed. Based upon computational experiments we suggest a value for $\alpha$ between 0.1 and 0.3.

b. Intermediate Hubs ($H^I$): Define $i$ as an intermediate hub if $i \in (I^{k-1} \cap H) \ni i \notin H^C, i \in Conv_{p^*}$. That is, these are hub locations (open or closed) in the previous consolidated region that also lie within the convex cone containing the national center and all potential hubs in the new region.

c. Non-critical Hubs ($H^N$): Defined as hubs in $I^k \cap H$ that do not belong to $H^C$ or $H^I$.

**STEP 5**: Reduced form of MIP-1

In this step we add constraints to MIP-1 based upon our classification of hubs:

a. Critical Hubs ($H^C$): Since the disposition of such a hub is more likely to change during consolidation, we impose no further restrictions on these.

b. Intermediate Hubs ($H^I$): For every intermediate hub, add constraints that maintain the same clinic assignments that it had in the solution (if it was open), i.e., for $j \in H^I$ add:

$$\sum_{r \in R} \sum_{f \in F} W_{jrf} = \sum_{r \in R} \sum_{f \in F} w^*_{jrf} \qquad (13)$$

and for all $i \in C$, $t \in T$

$$X_{ji} = x^*_{ji} \qquad (14)$$

$$U_{jit1} = u^*_{jit1} \qquad (15)$$

where $u^*_{jitf}$, $x^*_{ji}$, and $w^*_{jrf}$ are from the solution to the MIP defined on $G[I^{k-1}]$. Here (13) ensures that open intermediate hubs remain open and closed ones remain closed. Note that since such a hub can potentially supply other hubs, the required capacity of its own storage device and of the inbound transport device might increase, and the replenishment frequency at the hub might also change. Constraints (14) and (15) ensure the same flow into a clinic with the same device and replenishment frequency.

c. Non-critical Hubs ($H^N$): Add constraints for each open hub $j \in H^N$ to fix replenishment frequency, storage device, inbound and outbound volumes and vehicle types, and clinic assignment to be the same as they are in the solution to (*i*) the MIP defined on $G[I^{k-1}]$ if $j \in I^{k-1} \cap H$, or (*ii*) MIP-1 defined on $G[Q_{j^*}]$ if $j \in H_{j^*}$ and ensure that closed hubs remain closed. That is, for $j \in H^N$, $i \in C$, $l \in N \cup H$, $r \in R$, $t \in T$, $f \in F$, add:

$$X_{ji} = x^*_{ji} \qquad (16)$$

$$U_{jit1} = u^*_{jit1} \qquad (17)$$

$$W_{jrf} = w^*_{jrf} \qquad (18)$$

$$X_{lj} = x^*_{lj} \qquad (19)$$

$$U_{ljtf} = u^*_{ljtf} \qquad (20)$$

where $w^*_{jrf}$, $x^*_{ji}$, $u^*_{jitf}$, $x^*_{lj}$, $u^*_{ljtf}$ are values obtained from the solution to the MIP defined on $G[I^{k-1}]$ if $j \in I^{k-1} \cap H$, or the MIP on $G[Q_{j^*}]$ if $j \in H_{j^*}$.

Note that (18) maintains the open/closed status of a hub, (16) and (17) maintain the same flow into and the same transport device and replenishment frequency for each clinic served by a hub, while (19) and (20) do the same for the inbound flow into the hub (note that this last feature is different than with intermediate hubs).

**STEP 6:** Consolidation

With the constraints added in **STEP 5**, solve the MIP defined on subgraph $G[I^k]$. If $H' = \phi$, we have merged all hubs; stop and return the solution. Otherwise, delete all new constraints added in **STEP 5**, return to **STEP 3** to add a new region, and repeat the process at the next iteration.

**3.2 A refinement to Algorithm 1**

In **STEP 5**, there could potentially be hundreds of constraints added at each iteration. We can further manipulate the formulation at this step to obtain the same outcome but with fewer nodes in the graph. Instead of directly formulating the MIP on $G[I^k]$ with the constraints added in **STEP 5**, we could use information previously obtained from the solutions to problems defined on $G[I^{k-1}]$ and $G[Q_{p^*}]$ in order to restrict the problem size. Consider an open intermediate or non-critical hub $j$ that will be restricted to remain open at the next iteration along with the same clinic assignments. To reduce the number of nodes (and hence, the number of binary variables) we could collapse all clinics associated with the hub into a single dummy clinic $m$ with a demand equal to the sum of the demands at these clinics, locate it at the same location as the hub (so that $d_{jm} = 0$) and assign it to hub $j$. This ensures that the outflows to clinics served by $j$ are the same, so that the solution to the new problem will be the same as the one to the MIP on $G[I^k]$. The only difference is that in the modified problem the total transportation cost to the clinics served by $j$ is zero; however, we can simply add the true cost to the final value obtained by the new MIP.

More formally, consider a hub $j \in H^I \cup H^N$ that is open in the solution to the MIP on $G[I^{k-1}]$ or MIP-1 on $G[Q_{p^*}]$. Define a dummy clinic $m$ with demand $D_j^T$ equal to the total demand across all clinics served by hub $j$ in this solution.

$$D_j^T = \sum_{i \in c} \sum_{t \in T} b_i u^*_{jit1} \tag{21}$$

Also, define the new index set $C^-$ by removing from set $C$ the indices of all of the clinics serviced by hub $j$. Then we have the following proposition.

**PROPOSITION 1:** Given that hub $j \in H^I \cup H^N$ is open in the solution to $G[I^{k-1}]$ or $G[Q_{p^*}]$, MIP-1 with the additional constraints (14) and (15) for all $i \in C$, $t \in T$ is equivalent to MIP-1 with the following three additional constraints:

$$X_{jm} = D_j^T \tag{22}$$

$$\sum_{t \in T} U_{jmt1} = 1 \tag{23}$$

$$\sum_{l \in C^-} \sum_{t \in T} U_{jlt1} = 0 \tag{24}$$

Proof: See Appendix A.

By using Proposition 1, for every open intermediate or non-critical hub, we could replace the $\{|C| + |C| \times |T|\}$ constraints in (14) and (15) with the 3 constraints in (22), (23) and (24). Similarly, we could replace $\{|T| \times |C|\}$ binary variables associated with selecting devices used to send vaccines from the hub to all of its clinics, with $|T|$ binary variable associated with the dummy clinic at the hub and $|T| \times |C^-|$ binary variables associated with each of the clinics not consolidated into the dummy. This results in a reduction of $\{|T| \times (|C| - |C^-| - 1)\}$ in the number of binary variables. As a direct consequence of Proposition 1, we have the following:

**PROPOSITION 2:** For any hub $j \in H^I$ that is open in the solution to $G[I^{k-1}]$, MIP-1 with the constraints added in **STEP 5(b)** is equivalent to MIP-1 with constraints given by (22), (23) and (24) along with (13).

**PROPOSITION 3:** For any hub $j \in H^N$ that is open in the solution to $G[I^{k-1}]$ or $G[Q_{p^*}]$, MIP-1 with the constraints added in **STEP 5(c)** is equivalent to MIP-1 with constraints given by (22), (23) and (24), along with the constraints (18), (19) and (20) for $l \in N \cup H$, $r \in R$, $t \in T$, $f \in F$.

Note that if we use Propositions 2 and 3 to solve the modified formulation (as opposed to the one in **STEP 5**), we will need to add to the final objective value the following outbound transportation cost ($C_j^P$) for each hub $j$ that has its clinics consolidated:

$$C_j^P = \sum_{i \in C} \sum_{t \in T} 2c_{jit}^P g_1 d_{ji} u_{jit1}^* \tag{25}$$

Based on the preceding discussion, we have the following refined **Algorithm 1***:

**STEPS 1\* to 4\*:** Identical to **STEPS 1 to 4** in Algorithm 1.

**STEP 5\*:** Shrinkage

As in **STEP 5**, we first formulate MIP-1 for $G[I^k]$ but then add constraints and operations based on the category of the hub as follows:

a. Critical Hubs ($H^C$): No additional action or restrictions.

b. Intermediate Hubs ($H^I$): If hub $j \in H^I$ is open in the solution to $G[I^{k-1}]$, delete all clinics assigned to that hub, add a dummy clinic $m$ with demand $D_j^T$ computed via (21), and add the constraints given by (13), (22), (23) and (24) for that hub.

On the other hand, if hub $j$ is closed in the corresponding solution, add constraints to keep the hub closed:

$$\sum_{r \in R} \sum_{f \in F} W_{jrf} = 0$$

c. Non-critical Hubs ($H^N$): if a hub $j \in H^N$ is open in the solution to the problem on $G[I^{k-1}]$ or $G[Q_{p^*}]$, delete all clinics assigned to that hub, add a dummy clinic $m$ with demand $D_j^T$ computed via (21), and add the constraints given by (18), (19), (20), (22), (23) and (24) for that hub.

If hub $j$ is closed, add constraints to keep the hub closed:

$$\sum_{r \in R} \sum_{f \in F} W_{jrf} = 0$$

**STEP 6\*:** Consolidation

Identical to **STEP 6** except that we use (25) to add the total cost of additional transportation ($C_{Total}^*$) to the optimal value to account for the clinics deleted and consolidated in **STEP 5\***:

$$C_{Total}^* = \sum_{j \in H^I} C_j^P + \sum_{j \in H^N} C_j^P$$

## 4. Numerical experiments

We tested Algorithm 1* as well as a standard commercial solver on a suite of 43 different problems that are derived from the EPI networks in four different countries in sub-Saharan Africa. Due to data confidentiality issues, we denote these as Countries A, B, C and D. Several geographic and demographic characteristics of these four countries are shown in Table 2; for areas and population densities we have normalized the largest values to 1.0 and expressed the other values as respective fractions of these. As one can observe, there are significant differences in these. Countries A and B are relatively large but the population densities are relatively low. Most of the population in Country A is concentrated in a few regions with the remainder being sparsely distributed over the rest of the country in remote desert areas). In contrast, Countries C and D are smaller in area but densely populated and have many more existing vaccination facilities per km² of land.

**Table 2.** Characteristics

| Country | A | B | C | D |
|---|---|---|---|---|
| Area ($10^3$ km$^2$) | 0.99 | 1.00 | 0.09 | 0.45 |
| Population density [33] (per km$^2$) | 0.17 | 0.12 | 1.00 | 0.88 |
| No. of potential hubs ($h$) | 41 | 60 | 87 | 141 |
| Total no. of nodes ($n$+1) | 685 | 933 | 746 | 2875 |

We summarize detailed information on facilities, storage, transportation devices and vaccines in Tables 3 through 6 in Appendix B. Note that each country may have different transportation and storage devices to choose from and there can be significant differences in costs as well. For the problem instances that we considered, it happened that $|T|$=3 and $|R|$=4, although there were differences in the specific transport/storage device options in each country as shown in Tables 4 and 5. There are also minor differences in the EPI vaccine regimens within the countries. To obtain the total demand volume at each clinic we first estimate the number of newborns it must serve by multiplying the estimated population in the area served by the clinic and the corresponding national birth rate published by the World Bank. This is then multiplied by the number of required doses and the volume of each dose for each vaccine in the regimen, and adjusted upward to account for anticipated open-vial waste. Finally, the volumes are added across all vaccines.

We tested the algorithm using a computer with an Intel Core i5-6500 CPU and 3.20 GHz processor with 8.0 GB memory. For solving MIP-1 directly we used IBM ILOG CPLEX 12.6. Since none of the complete problems for any of the four countries could be directly solved, we started with smaller subproblems and worked on successively larger ones based on how a region is currently defined in the country. Detailed results for each of the four countries studied may be found in Tables 7 through 10 in Appendix C. Each entry in a table corresponds to a problem over a region, part of a region, or a set of regions in the corresponding country. The last row in each table denotes the problem with the full set of nodes across all regions of the country. We summarize the number of potential hubs, the total number of nodes, total number of binary variables and a characterization of the population density associated with each instance, in order to illustrate the diversity of the problems that we formulated. For the problems whose optimal solutions could be obtained via CPLEX, we report the percentage gap between the optimum cost and the objective value for the solution returned by Algorithm $1^*$. For a design problem such as this, computational times are obviously less relevant than being able to solve the problem; nevertheless, as a matter of reference we also list the solution times for MIP-1 using the commercial software (when an optimum solution is available) and for the solution found by Algorithm $1^*$.

As the tables show, the number of nodes, potential hubs, and binary variables in the largest problem that CPLEX could solve directly vary with each country. While problem size is certainly a factor, the geographical and population characteristics of the underlying network for a problem also play a role in determining whether it can be solved optimally. Based on extensive testing, it is our conclusion that (with a few exceptions) direct solution of MIP-1 using commercial software is feasible only in problems with fewer than approximately 200 nodes and 15 potential hubs, which is much smaller than the full network for almost all countries.

While the results indicate that the ability (and the effort) required to solve a problem optimally depends on the combination of factors listed in the tables, there is no systematic relationship with any one specific factor that could be established. However, as might be expected, the total number of binary variables seems significant. To further interpret the results we label our test problems as "large" or "small" using a cutoff of 20,000 binary variables. This leads to a total of 24 small and 19 large instances (including the full problem for each country) in our test set; the distribution of these labels for each country was different and depended on the specific characteristics of that country. The results show that while all the small instances could be solved optimally, MIP-1 corresponding to 15 of the 19 large instances could not be solved to optimality. In particular, countries C and D which have denser populations with more nodes per unit area proved harder to handle. Even for the instances that could be solved to optimality, the required effort can be inconsistent. For example, there are a couple of small instances (instances 8, 9 for Country A) that took a long time to solve, and while three of the four large instances that could be solved yielded solutions in reasonable times, one (instance 11, Country A) took over a week to solve. This was also the largest problem that we were able to solve to optimality. Also, in the case of the problems that could not be solved, there was no pattern to the integrality gap when the solver failed.

On the other hand, the disaggregation-and-merging approach of Algorithm $1^*$ was robust and able to generate solutions for every problem that we formulated (and in well under about 5 minutes in almost all cases; even the largest problem that we tested, with over 2 million variables, took only approximately 12 minutes). In the 28 instances where the optimal solution was available for comparison, Tables 7-10 show that Algorithm $1^*$ also converged to the optimal solution in 22 instances while finding a solution with a cost within 0.5% of the optimum value for five of the six remaining instances; our cost for the largest problem that could be solved optimally was 0.69% higher than the optimal cost. Moreover, even though the demographic characteristics also have an effect, Figure 3 shows that the computational effort appears to be approximately linear in the number of binary variables (the data point for the full Country D problem with over 2 million binary variables is omitted in the graph to maintain a better visual scale; the effort for this problem is actually proportionately smaller).

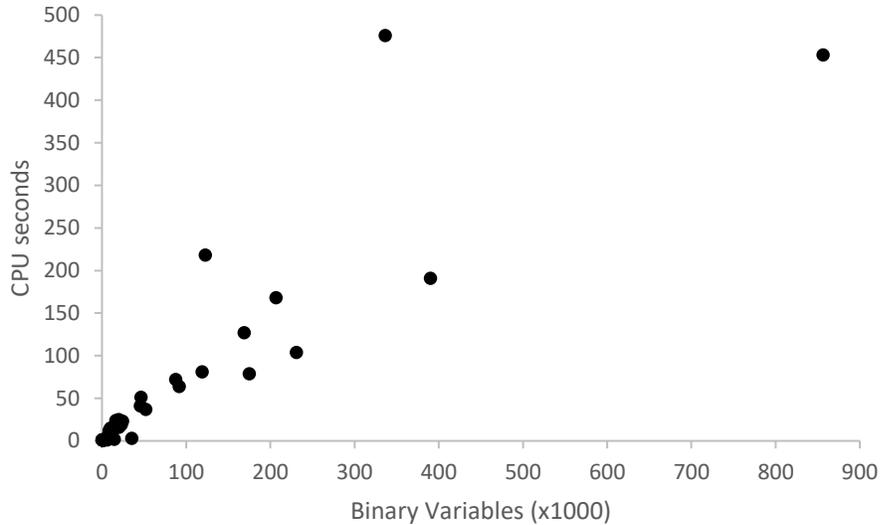

Figure 3: Computational times with Algorithm 1*

With the 15 problems for which the optimal solutions to MIP-1 are not available there was no way to compute the cost difference between the solution from Algorithm 1* and the optimum value. We could visually verify that the network structures generated were reasonable in all instances, and while there is no guarantee that they are optimal, they are certainly better than anything that could be derived by inspection or *ad hoc* methods. One comparison that we can do is to compare the current cost for the entire country against that of the solution generated by our algorithm (the last row in each of Tables 7-10), since this is an existing network. To compute the country-wide costs for the current structure, we used the same unit costs for facilities, transport and storage as those used in our numerical tests. The results are shown in Table 11 in Appendix C. Note that the total costs in both cases include identical operational costs for the national center as well as the clinics. The difference is that for the existing structure we have operational costs for regional and district centers and transportation costs from the national center to regions, regions to districts, and districts to clinics; while in our case, we have operational costs for the selected hubs and transport costs into the hubs and from the hubs to clinics (as captured by cost expression (1) in the formulation). Also note that we do not consider all the costs associated with the existing network in the comparison; rather we compare only the storage and transport costs currently incurred for EPI vaccines, with the corresponding costs in the redesigned network. The results indicate that even though we do not have a guarantee of optimality our distribution network produces overall savings ranging from approximately 6% to 27% for the four countries studied.

## 5. Summary and directions for future work

The WHO-EPI vaccine distribution network is of critical importance in low and middle-income countries, and designing it optimally can be of significant economic and social benefit to these countries. In this paper we present a general MIP formulation of the design problem that is applicable to any country. While this problem can be solved optimally when the network is small, it rapidly becomes intractable as the problem size grows and a different solution approach is needed to address the problem for an entire country. We present a novel MIP-based disaggregation-and-merging algorithm that is based on the simple observation that changes to the structure in a part of the network are unlikely to have a significant effect on the structure in other parts that are far away. The algorithm thus uses a divide-and-conquer approach to intelligently generate and solve a sequence of MIPs. Extensive tests based on real-world data derived from four different countries in sub-Saharan Africa show that it yields solutions that are optimal or within 0.5% of the best cost where optimality can be verified, and for large instances that are impossible to solve optimally, it is uniformly robust and yields good solutions in a few minutes.

There are several directions for future work. First, a possible limitation of our approach is the fact that it is designed for problems where it is reasonable to assume that structures for subnetworks that are physically distant will tend to be independent of each other. While this is generally true if costs are uniform, it might not be true in other networks. For example, it might be appropriate to locate the hub serving a portion of the network at a location that is far away if the associated costs there are significantly lower. It would be interesting to see what degradation is obtained in the results when our approach is applied to such networks. Second, one could consider and model uncertainty, which could be associated with both demand and supply. Finally, a related direction would be to develop more sophisticated vaccine inventory management policies than the current practice of using an across-the-board buffer of 25% with a fixed monthly/quarterly replenishment interval. From an implementation standpoint though, the challenge here is twofold. First, to quantify and evaluate stochasticity we would need much more data than is currently available.

Second, from a personnel standpoint, it would require far more sophistication in vaccine inventory management than is currently available in most LMICs.

## Acknowledgement

This work was partially supported by the National Science Foundation via Award No. CMII-1536430.

## Bibliography

[1]   World Health Organization. Immunization 2019. http://www.who.int/topics/immunization/en/ (accessed November 25, 2019).

[2]   World Health Organization. Immunization coverage 2018. http://www.who.int/news-room/fact-sheets/detail/immunization-coverage (accessed November 25, 2019).

[3]   Gavi. Gavi Progress Reports. Glob Vaccine Alliance 2017. https://www.gavi.org/library/publications/gavi-progress-reports/gavi-progress-report-2017/ .

[4]   Yadav P, Lydon P, Oswald J, Dicko M, Zaffran M. Integration of vaccine supply chains with other health commodity supply chains: A framework for decision making. Vaccine 2014;32:6725–32. https://doi.org/10.1016/j.vaccine.2014.10.001.

[5]   Zaffran M. Vaccine transport and storage: environmental challenges. Dev Biol Stand 1996;87:9—17.

[6]   Bland J, Clements CJ. Protecting the world's children: the story of WHO's immunization programme. World Heal. forum 1998; 19, 1998, p. 162–73.

[7]   Gavi. Our impact. Gavi, the Vaccine Alliance n.d. https://www.gavi.org/programmes-and-impact/our-impact (accessed November 25, 2019).

[8]   World Health Organization. The Expanded Programme on Immunization 2013. http://www.who.int/immunization/programmes_systems/supply_chain/benefits_of_immunization/en/ (accessed November 25, 2019).

[9]   World Health Organization. Prequalified Devices and Equipment n.d. http://apps.who.int/immunization_standards/vaccine_quality/pqs_catalogue/categorylist.aspx?cat_type=device.

[10]  Şahin G, Süral H. A review of hierarchical facility location models. Comput Oper Res 2007;34:2310–31. https://doi.org/10.1016/j.cor.2005.09.005.

[11]  Melo MT, Nickel S, Saldanha-da-Gama F. Facility location and supply chain management

- A review. Eur J Oper Res 2009;196:401–12. https://doi.org/10.1016/j.ejor.2008.05.007.

[12] Mirchandani PB. The p-median problem and generalizations. In: Mirchandani P, Francis R, editors. Discret. Locat. Theory, New York: John Wiley and Sons Inc; 1990, p. 55–117.

[13] Geoffrion A. M., Graves G. W. Multicommodity Distribution System Design by Benders Decomposition. Manage Sci 1974;20:822–44.

[14] Wu TH, Low C, Bai JW. Heuristic solutions to multi-depot location-routing problems. Comput Oper Res 2002;29:1393–415. https://doi.org/10.1016/S0305-0548(01)00038-7.

[15] Shen ZM, Coullard C, Daskin MS. A Joint Location-Inventory Model. Transp Sci 2003;37:1–121.

[16] Snyder L V., Daskin MS. Reliability models for facility location: The expected failure cost case. Transp Sci 2005;39:400–16. https://doi.org/10.1287/trsc.1040.0107.

[17] Daskin MS, Snyder L V., Berger RT. Facility location in supply chain design. In: Langevin A, Riopel D, editors. Logist. Syst. Des. Optim., Springer US; 2005, p. 39–65. https://doi.org/10.1007/0-387-24977-X_2.

[18] Klose A, Drexl A. Facility location models for distribution system design. Eur J Oper Res 2005;162:4–29. https://doi.org/10.1016/j.ejor.2003.10.031.

[19] Farahani RZ, Hekmatfar M, Arabani AB, Nikbakhsh E. Hub location problems: A review of models, classification, solution techniques, and applications. Comput Ind Eng 2013;64:1096–109. https://doi.org/10.1016/j.cie.2013.01.012.

[20] Campbell JF, O'Kelly ME. Twenty-five years of hub location research. Transp Sci 2012;46:153–69. https://doi.org/10.1287/trsc.1120.0410.

[21] da Graça Costa M, Captivo ME, Clímaco J. Capacitated single allocation hub location problem-A bi-criteria approach. Comput Oper Res 2008;35:3671–95. https://doi.org/10.1016/j.cor.2007.04.005.

[22] Correia I, Nickel S, Saldanha-da-Gama F. Single-assignment hub location problems with multiple capacity levels. Transp Res Part B Methodol 2010;44:1047–66. https://doi.org/10.1016/j.trb.2009.12.016.

[23] Kratica J, Milanović M, Stanimirović Z, Tošić D. An evolutionary-based approach for solving a capacitated hub location problem. Appl. Soft Comput. J., vol. 11, 2011, p. 1858–66. https://doi.org/10.1016/j.asoc.2010.05.035.

[24] De Camargo RS, Miranda G. Single allocation hub location problem under congestion: Network owner and user perspectives. Expert Syst Appl 2012;39:3385–91. https://doi.org/10.1016/j.eswa.2011.09.026.

[25] Taghipourian F, Mahdavi I, Mahdavi-Amiri N, Makui A. A fuzzy programming approach for dynamic virtual hub location problem. Appl Math Model 2012;36:3257–70.

https://doi.org/10.1016/j.apm.2011.10.016.

[26] Ernst AT, Krishnamoorthy M. Solution algorithms for the capacitated single allocation hub location problem. Ann Oper Res 1999;86:141–59. https://doi.org/10.1023/A:1018994432663.

[27] Contreras I, Díaz JA, Fernández E. Lagrangean relaxation for the capacitated hub location problem with single assignment. OR Spectr 2009;31:483–505. https://doi.org/10.1007/s00291-008-0159-y.

[28] De Camargo RS, De Miranda G, Ferreira RPM. A hybrid Outer-Approximation/Benders Decomposition algorithm for the single allocation hub location problem under congestion. Oper Res Lett 2011;39:329–37. https://doi.org/10.1016/j.orl.2011.06.015.

[29] Chen JF. A heuristic for the capacitated single allocation hub location problem. Lect. Notes Electr. Eng., vol. 5 LNEE, 2008, p. 185–96. https://doi.org/10.1007/978-0-387-74905-1_14.

[30] Chen SI, Norman BA, Rajgopal J, Assi TM, Lee BY, Brown ST. A planning model for the WHO-EPI vaccine distribution network in developing countries. IIE Trans (Institute Ind Eng 2014;46:853–65. https://doi.org/10.1080/0740817X.2013.813094.

[31] Lim J. Improving the design and operation of WHO-EPI vaccine distribution networks. University of Pittsburgh, 2016.

[32] Charu C. Aggarwal JH. Data Mining: The Textbook. Springer Int Publ 2015:746. https://doi.org/10.1007/978-3-319-14142-8.

[33] United Nations. World Population Prospects 2017. https://population.un.org/wpp/DataQuery/ (accessed November 26, 2019).

# Appendix A: Proof of Proposition 1

Since hub $j$ is open in the solution to $G[I^{k-1}]$ or $G[Q_{j^*}]$, we have

$$\sum_{r \in R} \sum_{f \in F} W_{jrf} = 1.$$

First, suppose that for this $j$ and all $i \in C$, $t \in T$ we add the constraints given by (14) and (15). This is equivalent to partially fixing the network structure. Specifically, we fix the clinic assignments for hub $j$. With constraint (7) we have for all $i \in C$ that are served by hub $j$

$$X_{ji} = x^*_{ji} = b_i.$$

Therefore, the total volume that goes out of hub $j$ and goes into all of its clinics is also fixed:

$$\sum_{i \in C} X_{ji} = \sum_{i \in C} X_{ji} \sum_{t \in T} U_{jit1} = \sum_{i \in C} \sum_{t \in T} X_{ji} U_{jit1} = \sum_{i \in C} \sum_{t \in T} b_i u^*_{jit1} = D_j^T.$$

Note that the second and third equalities above hold because of constraint (2).

Now, suppose instead that we delete all clinics assigned to $j$ to obtain the index set $C^-$ and add a dummy clinic $m$ with demand given by (21), then add constraints (22), (23) and (24) to MIP-1. From these constraints and constraint (6) of MIP-1, once again the total volume that goes out of hub $j$ is fixed, i.e., we have:

$$\sum_{l \in C} X_{jl} = X_{jm} = D_j^T.$$

Since we are not altering any other constraints, this is equivalent to the first case with fixed clinic assignments and the only difference is that the same total outflow is sent to a single clinic. Thus the optimal solutions with either approach are identical. □

# Appendix B: Input Data

**Table 3.** Facility Cost ($/year)

| Facility type | Country A | Country B | Country C | Country D |
|---|---|---|---|---|
| National | 40,000 | 14,870 | 52,500 | 158,191 |
| Hub | 4,500 | 450 | 2,389 | 20,992 |
| Clinic | 800 | 150 | 140 | 1,825 |

**Table 4.** Storage Details

| Country | Device Type | Capacity (L) | Cost ($/year) |
|---|---|---|---|
| A | Cold Room 1 | 18,000 | 8,116 |
|   | Cold Room 2 | 1,200 | 1,200 |
|   | Refrigerator 1 | 700 | 900 |
|   | Refrigerator 2 | 340 | 610 |
| B | Cold Room 1 | 5,000 | 17,534 |
|   | Cold Room 2 | 1,500 | 1,800 |
|   | Refrigerator 1 | 700 | 900 |
|   | Refrigerator 2 | 504 | 624 |
| C | Cold Room | 1,500 | 1,500 |
|   | Refrigerator 1 | 340 | 650 |
|   | Refrigerator 2 | 200 | 550 |
|   | Refrigerator 3 | 53 | 462 |
| D | Cold Room | 5,000 | 17,534 |
|   | Refrigerator 1 | 504 | 624 |
|   | Refrigerator 2 | 340 | 510 |
|   | Refrigerator 3 | 84 | 394 |

**Table 5.** Transportation Details

| Country | Vehicle Type | Capacity (L) | Cost ($/km) |
|---------|--------------|--------------|-------------|
| A | Cold truck | 9,293 | 0.97 |
| | 4×4 truck | 172 | 0.54 |
| | Motorbike | 5 | 0.23 |
| B | Cold truck | 9,500 | 0.78 |
| | 4×4 truck | 308.44 | 0.51 |
| | Motorbike | 3.4 | 0.1 |
| C | Truck 1 | 331.2 | 1.4 |
| | Truck 2 | 110.4 | 0.4667 |
| | Motorbike | 3 | 0.13 |
| D | Cold truck | 15,000 | 1.12 |
| | 4×4 truck | 82.8 | 0.38 |
| | Motorbike | 3 | 0.12 |

**Table 6.** Vaccine Details

| Country | Name | Pharmaceutical Form | Open-vial waste | Dose Volume (cc) | Required Dosage |
|---|---|---|---|---|---|
| A | Tuberculosis | Lyophilized | 0.5 | 1.2 | 1 |
|   | Tetanus Toxoid | Liquid | 0 | 3 | 3 |
|   | Measles | Lyophilized | 0.4 | 2.1 | 1 |
|   | Oral Polio | Liquid | 0 | 1 | 4 |
|   | Yellow Fever | Lyophilized | 0.4 | 2.5 | 1 |
|   | DTC-HepB-Hib | Liquid | 0 | 16.8 | 3 |
|   | Rotavirus | Liquid | 0 | 45.9 | 3 |
|   | PCV13 | Liquid | 0 | 12 | 3 |
| B | Tuberculosis | Lyophilized | 0.5 | 1.2 | 1 |
|   | Tetanus Toxoid | Liquid | 0.15 | 3 | 2 |
|   | Measles | Lyophilized | 0.45 | 3.5 | 1 |
|   | Oral Polio | Liquid | 0.17 | 1 | 4 |
|   | Yellow Fever | Lyophilized | 0.45 | 2.5 | 1 |
|   | DTC-HepB-Hib | Liquid | 0.1 | 11 | 3 |
|   | Rotavirus | Liquid | 0 | 17.1 | 2 |
|   | PCV13 | Liquid | 0.05 | 12 | 3 |
| C | Tetanus Toxoid | Liquid | 0.05 | 3 | 2 |
|   | Measles | Lyophilized | 0.45 | 3.5 | 1 |
|   | Oral Polio | Liquid | 0.17 | 1 | 4 |
|   | Yellow Fever | Lyophilized | 0.1 | 2.5 | 1 |
|   | BCG | Lyophilized | 0.5 | 1.2 | 1 |
|   | Pentavalent | Liquid | 0.15 | 5.3 | 3 |
|   | Pentavalent | Liquid | 0.45 | 12.9 | 3 |
| D | Tetanus Toxoid | Liquid | 0.1 | 2.5 | 2 |
|   | Measles | Lyophilized | 0.45 | 3.5 | 1 |
|   | Oral Polio | Liquid | 0.17 | 2 | 4 |
|   | Yellow Fever | Lyophilized | 0.05 | 2.46 | 1 |
|   | PCV10 | Liquid | 0.45 | 4.8 | 3 |
|   | BCG | Lyophilized | 0.5 | 1.2 | 1 |
|   | Pentavalent | Liquid | 0.15 | 9.7 | 3 |

# Appendix C: Computational Results

**Table 7.** Computational results for Country A

| No. | Hubs | Nodes | Binary Variables | Size Label | Population Density | Gap | CPU Times MIP-1 | CPU Times Algorithm 1[*] |
|---|---|---|---|---|---|---|---|---|
| 1 | 1 | 10 | 68 | small | sparse | 0% | <1s | <1s |
| 2 | 2 | 49 | 604 | small | sparse | 0% | <1s | <1s |
| 3 | 4 | 48 | 1,184 | small | moderate | 0% | <1s | <1s |
| 4 | 5 | 77 | 2,350 | small | moderate | 0% | 1.6s | <1s |
| 5 | 7 | 99 | 4,214 | small | dense | 0% | 4.4s | 2s |
| 6 | 8 | 206 | 9,952 | small | dense | 0% | ~10h | 15s |
| 7 | 14 | 148 | 12,544 | small | moderate | 0% | 103s | 14s |
| 8 | 13 | 210 | 16,484 | small | moderate | 0% | ~1d | 24s |
| 9 | 14 | 235 | 19,852 | small | dense | 0% | ~2d | 25s |
| 10 | 19 | 176 | 20,216 | *large* | moderate | 0% | 4,649s | 16s |
| 11 | 20 | 384 | 46,240 | *large* | dense | 0.29% | ~1.1w | 51s |
| 12 | 33 | 599 | 118,866 | *large* | moderate | - | - | 81s |
| Full | 41 | 685 | 168,838 | *large* | moderate | - | - | 127s |

**Table 8.** Computational results for Country B

| No. | Hubs | Nodes | Binary Variables | Size Label | Population Density | Gap | CPU Times MIP-1 | CPU Times Algorithm 1[*] |
|---|---|---|---|---|---|---|---|---|
| 1 | 8 | 56 | 2,752 | small | sparse | 0% | 16s | 1s |
| 2 | 14 | 101 | 8,596 | small | sparse | 0% | 119s | 12s |
| 3 | 12 | 128 | 9,312 | small | dense | 0% | 116s | 13s |
| 4 | 16 | 162 | 15,680 | small | dense | 0% | 1,304s | 15s |
| 5 | 28 | 271 | 45,752 | *large* | dense+moderate | - | - | 41s |
| 6 | 41 | 372 | 91,840 | *large* | moderate | - | - | 64s |
| 7 | 57 | 510 | 174,876 | *large* | moderate | - | - | 79s |
| 8 | 65 | 591 | 231,010 | *large* | moderate+sparse | - | - | 104s |
| Full | 87 | 746 | 390,108 | *large* | moderate+sparse | - | - | 191s |

**Table 9.** Computational results for Country C

| No. | Hubs | Nodes | Binary Variables | Size Label | Population Density | Gap | CPU Times MIP-1 | Algorithm 1* |
|-----|------|-------|------------------|------------|--------------------|-----|-----------------|--------------|
| 1 | 1 | 18 | 116 | small | moderate | 0% | <1s | <1s |
| 2 | 2 | 11 | 148 | small | sparse | 0% | <1s | <1s |
| 3 | 3 | 22 | 420 | small | sparse | 0% | 2s | <1s |
| 4 | 3 | 39 | 726 | small | moderate | 0% | 2s | <1s |
| 5 | 4 | 44 | 1,088 | small | sparse | 0% | 3s | <1s |
| 6 | 4 | 55 | 1,352 | small | moderate | 0% | 4s | <1s |
| 7 | 4 | 64 | 1,568 | small | moderate | 0% | 7s | <1s |
| 8 | 4 | 65 | 1,592 | small | dense | 0% | 8s | <1s |
| 9 | 11 | 96 | 6,424 | small | moderate | 0.07% | 10s | <1s |
| 10 | 17 | 141 | 14,518 | small | moderate | 0.15% | 79s | 2s |
| 11 | 20 | 295 | 35,560 | *large* | moderate | 0.42% | 387s | 3s |
| 12 | 26 | 333 | 52,156 | *large* | sparse+moderate | 0.69% | 2,748s | 37s |
| 13 | 34 | 601 | 122,876 | *large* | moderate+dense | - | - | 218s |
| Full | 60 | 933 | 336,360 | *large* | sparse+dense | - | - | 476s |

**Table 10.** Computational results for Country D

| No. | Hubs | Nodes | Binary Variables | Size Label | Population Density | Gap | CPU Times MIP-1 | Algorithm 1* |
|-----|------|-------|------------------|------------|--------------------|-----|-----------------|--------------|
| 1 | 10 | 117 | 7,100 | small | moderate | 0.25% | 146s | 6s |
| 2 | 14 | 270 | 22,792 | *large* | dense | - | - | 19s |
| 3 | 11 | 366 | 24,244 | *large* | dense | - | - | 23s |
| 4 | 27 | 540 | 87,696 | *large* | dense | - | - | 72s |
| 5 | 38 | 906 | 206,872 | *large* | dense | - | - | 168s |
| 6 | 83 | 1,718 | 856,228 | *large* | dense+moderate | - | - | 453s |
| Full | 141 | 2,875 | 2,433,378 | *large* | moderate | - | - | 713s |

**Table 11.** Network cost for Country A, B, C, D

| Country | A | B | C | D |
|---------|---|---|---|---|
| Original Cost ($) | 2,453,690 | 791,164 | 5,239,822 | 8,674,722 |
| Optimized Cost ($) | 1,844,129 | 743,903 | 3,819,622 | 7,869,399 |
| Savings | 24.84% | 5.97% | 27.10% | 9.28% |